\documentclass[lettersize,journal]{IEEEtran}
\usepackage[T1]{fontenc}
\usepackage[utf8]{inputenc}

\usepackage{amsmath,amsfonts}
\usepackage[noend,ruled,vlined]{algorithm2e}
\usepackage[noend]{algpseudocode}
\usepackage{array}
\usepackage[caption=false,font=normalsize,labelfont=sf,textfont=sf]{subfig}
\usepackage{textcomp}
\usepackage{stfloats}
\usepackage{url}
\usepackage{verbatim}
\usepackage{graphicx}
\usepackage{cite}
\usepackage{dsfont}
\usepackage[hidelinks]{hyperref}
\usepackage{mathtools}
\usepackage{amssymb}
\usepackage{booktabs}
\usepackage{cleveref}
\usepackage[numbers]{natbib}
\usepackage{xcolor}
\usepackage{authblk}

\newcommand{\argmin}[1]{\underset{#1}{\text{argmin}}\,}

\begin{document}

\title{National-scale bi-directional EV fleet control for ancillary service provision}
\author[1,2]{L. Nespoli\thanks{lorenzo.nespoli@supsi.ch}}
\author[3]{N. Wiedemann} 
\author[3]{ E. Suel} 
\author[3]{Y. Xin}
\author[3]{M. Raubal} 
\author[1]{V. Medici} 
\affil[1]{SUPSI, Mendrisio, Switzerland}
\affil[2]{Hive Power SA, Suglio, Switzerland}
\affil[3]{ETHZ, Zurich, Switzerland}
	



\maketitle

\begin{abstract}
Deploying real-time control on large-scale fleets of electric vehicles (EVs) is becoming pivotal as the share of EVs over internal combustion engine vehicles increases. In this paper, we present a Vehicle-to-Grid (V2G) algorithm to simultaneously schedule thousands of EVs charging and discharging operations, that can be used to provide ancillary services. To achieve scalability, the monolithic problem is decomposed using the alternating direction method of multipliers (ADMM). Furthermore, we propose a method to handle bilinear constraints of the original problem inside the ADMM iterations, which changes the problem class from Mixed-Integer Quadratic Program (MIQP) to Quadratic Program (QP), allowing for a substantial computational speed up. We test the algorithm using real data from the largest carsharing company in Switzerland and show how our formulation can be used to retrieve flexibility boundaries for the EV fleet. Our work thus enables fleet operators to make informed bids on ancillary services provision, thereby facilitating the integration of electric vehicles.
\end{abstract}


\section{Introduction}

\subsection{Background and motivation}

Public authorities and the private sector face many challenges in transforming industries and infrastructure to meet sustainability goals. A key factor is the successful integration of renewable energies such as solar or wind power, which however poses difficulties to the power system 
due to the increased fluctuations in supply from renewable energy sources. At the same time, an increasing number of electric vehicles pose an additional burden on the grid~\cite{iae}. Both challenges inspired the development of smart charging or V2G technologies, where the charging flexibility of EVs are exploited as buffer storage to the power system. Smart charging and V2G were shown to have high potential benefits for peak load shaving~\cite{xu2018planning, crozier2020opportunity, kempton2005vehicle}, 
supporting the integration of renewable energies~\cite{martin2022using} while offering additional revenues to vehicle owners~\cite{kara_estimating_2015}. 

Although smart charging and V2G have been studied for years~\citep{garcia-villalobos_plug-electric_2014, tan_integration_2016}, they remain difficult to implement in practice for the following reasons: 1) they require control over a sufficiently large fleet of EVs, 2) they imply complex dispatching problems, and 3) they involve trading between the power system and the vehicle fleet operators. A major opportunity is the application of V2G for large-scale car sharing systems~\citep{fournier2014carsharing}, since they can centrally manage large and significant resources for V2G operations. In contrast to the share of EVs on the private vehicle market (8\% global sales share\footnote{https://www.ev-volumes.com}), the share of EVs in car sharing systems is already high, with more than 66\% of car sharing services offering fully or partially electric fleets~\cite{shaheen2020innovative}. V2G may afford additional revenues to car sharing operators, but at the same time requires careful dispatching to minimize the negative impact on car availability for mobility purposes.

Here, we propose an optimization approach for V2G operations that scales to a large fleet of EVs. Specifically, we first provide a monolithic formulation for optimizing charging schedules, and further develop relaxations that allow to decompose the problem by aggregated vehicle hubs such as car sharing stations. Our experiments demonstrate a strong improvement in runtime using our approach, enabling its application on a large-scale vehicle fleet.
Furthermore, the optimization framework is tested on a new dataset from a car sharing operator in Switzerland. It is shown that our method scales to a fleet of 1440 electric vehicles in feasible runtime and can be employed to decrease energy costs while providing different kinds of grid services. Our optimization approach is therefore not only relevant for car sharing services but may in general support in controlling V2G fleet operations.

\subsection{Literature review and previous works}

An increasing number of works is tackling the problem of charging schedule optimization in the context of car sharing; \citet{xu_electric_2021} optimize charging times in a MINLP problem targeted at determining the fleet size of a car sharing system. \citet{he2021charging} optimize the charging station setup and schedule for a car sharing fleet and provide interesting insights on the best decisions on charging station placement and minimum State of charge (SOC). Similarly, \cite{biondi2016optimal} formulate a two-step optimization problem in order to reduce the charging prices in a shared system, while retaining user satisfaction.
Only some research has focused on large-scale, national level optimization of V2G, since this is a more challenging problem if realistic constraints are considered. Furthermore, the typical scale of pilot projects in this context is small: in \cite{raviUtilizationElectricVehicles2022} the authors reviewed 54 pilot projects using EVs for providing grid services, reporting an average number of 26 EVs per pilot. In \cite{maEfficientDecentralizedCoordination2016} a decentralized algorithm to optimize the charge (but not discharge) of 5000 EVs was presented. In \cite{yiHighlyEfficientControl2020} the authors present a rule-base two-stage hierarchical approach to coordinate charging operations of thousands of EVs. While this research only considers smart charging and not V2G, \cite{caggiani_static_2021} also include the possibility of V2G in the relocation-optimization of one-way car sharing. In \cite{zhongCoordinatedControlLargescale2014} the authors coordinated 500 EVs to achieve frequency regulation using a rule-based control in a V2G setting. \cite{zhang2021values} regard the problem that is most closely related to our formulation, namely V2G strategies for car sharing, and they propose a two-stage stochastic optimization employing a 24 hours receding horizon approach solved with a resolution of 15 minutes. They show that keeping integer variables lead to infeasible solution times (greater than 32 hours in their case), and propose to both relax all integer variables to continuous one and use decomposition techniques in order to speed it up. However, they do not provide a scalability analysis of their algorithms, nor mention the number of considered EVs.
In contrast to optimal control methods, others propose data-driven optimization with learning methods. For example, \cite{valogianni2013smart, wan2018model, tuchnitz2021development,dang_q-learning_2019,li2019constrained} train a reinforcement learning (RL) agent to decide on charging behavior.
%
%
However, these methods are usually focused on finding decision policies for single EVs, since finding the optimal joint actions for a fleet of EVs, which is the focus of our work, is a much more challenging task, in general requiring a multi-agent RL strategy, which usually involve to optimize over a large decision space. Authors in \cite{sadeghianpourhamami2019definition} propose RL for guiding charging decisions for a whole vehicle \textit{fleet} at once by reducing the action space by pooling EVs with similar energy requests; however, this was done not considering external inputs such as an aggregated profile, and disregarding V2G.

\section{Problem definition and formulation}
In the following we start describing a generic formulation needed to effectively synchronize the EV fleet charging and discharging operations, and later explain how relaxing some conditions can lower the overall computational complexity. 
The common setting for all the problem formulations is the following: a car sharing provider operating a stationary fleet (as opposed to free-floating) is willing to jointly optimize all its EVs' operations in order to reduce its own operating costs, whether by optimizing for a dynamic price, increasing its own self-consumption if local PV generation is present, or by providing services to the electric grid. 
Furthermore, the provider knows at least an approximated schedule of the future EV locations, in terms of their presence at a given charging station and driven mileage for the next control horizon. This can be realistically achieved using information from booking apps and by modeling historical data. Based on these assumptions we can estimate  the lower bounds for the EVs' battery energy constraints needed to satisfy all their foreseen mobility demand, as we will show in section \ref{sec:soc_modeling}. These time series are required to formulate the optimal control problem, as explained in the following section.

\subsection{Monolithic formulations}
Given a control horizon of $T$ steps, $n_s$ stations, each station hosting $n_{v, s}$ vehicles, and called $\mathcal{T}$ and $\mathcal{S}$ the sets of times and stations, the monolithic problem can be described as:   
\begin{align}
&u^{*} =  \argmin{\mathcal{X}}{F(u) + Q(x)} \label{eq:battery_of}\\ 
&x_{t+1, v} = A_v x_{t, v} + B_v u_{t, v} -\Delta e_{t, v} \quad  \forall t \in \mathcal{T}, v \in \mathcal{V} \label{eq:state_dyn}\\
&u \succcurlyeq 0 \label{eq:u_constr}\\
&u_c \preccurlyeq x_c u_{c, max}^T \qquad u_d \preccurlyeq (1-x_c) u_{d, max}^T\label{eq:u_bilin} \\
&u_c \preccurlyeq c u_{c, max}^T \qquad u_d \preccurlyeq c u_{d, max}^T\label{eq:u_loc} \\
&\sum_{v \in \mathcal{V}_{t, s}} u_{c, t, v} - u_{d, t, v} \in \mathcal{U}_s  \quad  \forall t \in \mathcal{T}, s \in \mathcal{S}  \label{eq:s_constr_u}\\
&\sum_{v \in \mathcal{V}_{t, s}} c_{t, v}  \leq  n_{max, s} \quad     \forall t \in \mathcal{T}, s \in \mathcal{S} \label{eq:s_constr_c}
\end{align}
where $x \in \mathds{R}^{T \times \sum_s n_{v,s}}$ is the matrix containing the battery state for all the EVs in kWh. For sake of clarity, table \ref{tab1} reports all the parameters and optimization variables $\mathcal{X}$ of the problem with associated dimensions and domains.

\begin{table}[ht]
\begin{tabular}{@{}llll@{}}
\toprule
Name                                & Type      & Dim.                 & Description                                    \\\midrule
$c$                                   & var  & $\mathds{Z}^{T \times n_v}_{\{0,1 \}}$ & ev connected to charger                    \\
$x_{c}$                         & var & $\mathds{Z}^{T \times n_v}_{\{0, 1\}}$              & plug state \\
$x$                                   & var  & $\mathds{R}^{T+1 \times n_{v}}$        & batteries state {[}kWh{]}                        \\
$u_c, u_d$                                    & var  & $\mathds{R}^{T \times n_{v}}$ & charging / discharging power {[}kW{]}        \\
$y$                                   & var  & $\mathds{R}^{T}$          & energy costs {[}£{]}                           \\
$r$                                   & par  & $\mathds{R}^{T}$          & reference profile                           \\
$l$                                   & par  & $\mathds{Z}^{T \times n_{v}}$          & location matrix                           \\
$n_{max, s}$                                   & par  & $\mathds{Z}^{n_{s}}$          & stations' chargers                            \\
$p_{s, max}$                                   & par  & $\mathds{R}^{n_{s}}$          & stations' max power                           \\
$p_{buy}$, $p_{sell}$             & par & $\mathds{R}^T$            & buying and selling prices {[}£/kWh{]}                       \\
$e$                          & par & $\mathds{R}^{T\times n_v}$            & energy constraint matrix {[}kWh{]}   \\
$\Delta e$                          & par & $\mathds{R}^{T\times n_v}$            & $\Delta$ energy at arrival {[}kWh{]}   \\
$\hat{p}$                                & par & $\mathds{R}^T$            & forecasted station power    {[}kW{]}              \\
$\hat{p}_{pv}$                                   & par  & $\mathds{R}^{T}$          & forecasted PV profile                           \\
$x_{start}$                        & par & $\mathds{R}^{n_v}$              & initial battery state {[}kWh{]}                \\

$x_{min}, x_{max}$                         & par & $\mathds{R}^{n_v}$              & capacity limits [kWh] \\
$u_{d, min}, u_{d, max}$                         & par & $\mathds{R}^{n_v \times 2}$              & discharging limits [kW] \\
$u_{c, min}, u_{c, max}$                         & par & $\mathds{R}^{n_v \times 2}$              & charging limits [kW] \\\bottomrule
\\
\end{tabular}
\caption{Variables, parameters and constants of the EV optimization problem.}\label{tab1}
\end{table}
Here $F(u): \mathds{R}^{T \times \sum_s n_{v,s}} \rightarrow \mathds{R}$ and $Q(x): \mathds{R}^{T \times \sum_s n_{v,s}} \rightarrow \mathds{R}$ are two scalar convex functions. In particular $F(u)$ is a cost function associated with the charging and discharging actions of the EVs and depends on the specific business model and will be further specified in section \ref{sec:decomposition}. We now explain in detail the problem constraints. Equation \ref{eq:state_dyn} describes the EVs dynamic equation, taking into account self-discharge and asymmetric charging and discharging efficiencies encoded in the $A_v \in \mathds{R}$ and $B_v \in \mathds{R}^2$ discrete dynamics matrices, obtained by the continuous one through exact discretization \cite{shieh_discrete-continuous_1980}:
\begin{equation}
\begin{array}{l}
A=e^{A_c d t} \\
B=A_c^{-1}\left(A_{d}-I\right) B_c
\end{array}
\end{equation}
where $A_c=\frac{1}{\eta_{sd}}$ and $B_c=[\eta_{ch}, \frac{1}{\eta_{ds}}]$, and $\eta_{sd}$, $\eta_{ch}$ and $\eta_{ds}$ are the characteristic self-discharge constant, charge and discharge efficiencies, respectively. Since $B_c$ defines an asymmetric behaviour in charging and discharging (even with equal charging/discharging coefficients), solving the battery scheduling requires to use two different variables for the charging and discharging powers for each EV. These are concatenated and denoted as a whole as $u = [u_c, u_d] $, where $u_d, u_c \in \mathds{R}^{T, n_v}$ are charging and discharging operations for all the EVs in kW. $\Delta e  \in \mathds{R}^{T, n_v}$ is the (sparse) matrix containing the energy lost during the last EV trip, defined as:
\begin{equation}\label{eq:delta_e}
\Delta e_{t, v} = 
\begin{cases}
    e_{t_d(t), v} \quad \text{if} \quad \Delta_t l_{t, v} > 0\\ 
    0 \quad \text{otherwise}    
\end{cases}
\end{equation}
where the first condition in equation \eqref{eq:delta_e} designs times in which the location matrix has a positive discrete derivative, that is, when the $v_{th}$ EV connects to a charging station. Here $e \in \mathds{R}^{T \times n_v}$ is the (sparse) energy constraint matrix, containing the energy that the EVs require at departure times, while $t_d(t)$ is the last departure time seen at step $t$. In other words, the minimum energies required at departure times and encoded in $e$ are equal to the energy drops $\Delta e_{t, v}$ needed to be reintegrated at next arrival time.  
The energy requirements stored in $e$ are assumed to be known at solution time for the next solution horizon, and they are estimated starting by the total driven km for the last trip, as explained in section \ref{sec:soc_modeling}.
Since it is not always possible to guarantee that all the EVs satisfy the energy requirements stored in $e$ at departure time, state constraints on the EVs SOC are taken into account as a threshold soft constraints encoded in $Q(x)$:
\begin{equation}\label{eq:soft_constr}
    Q(x) = k \Vert\text{max}(e - x, 0)\Vert_2^2
\end{equation}
where $k$ is a large constant, which allows to retrieve feasible solutions even if some EVs are not fully charged.
Equation \eqref{eq:u_constr} states that charging and discharging variables $u_c$ and $u_d$ are positive quantities. Equation \eqref{eq:u_bilin} makes use of the binary variable $x_c$, which indicates whether a given EV is charging, to encode the bilinear constraint $u_c\odot u_d = 0$, where $\odot$ is the Hadamard product; this encodes the fact that each EV cannot charge and discharge simultaneously. It must be noted that this condition is sometimes naturally satisfied by the problem, depending on the objective function $F(u)$, as shown for example in \cite{garifi_control_2019}. However, this is not always guaranteed; for example if we want to implement peak shaving in the presence of PV power plants. In this case EVs could occasionally decide to both charge and discharge and exploit the round-trip efficiency to dissipate more power and perform valley filling when the overall station network is a net energy producer. The same reasoning can be applied to quadratic profile tracking, as in the case of tracking a given power profile for providing services to the grid. 
In equation \eqref{eq:u_loc}, the binary variable $c \in \mathds{R}^{T \times n_v}$ is used to enforce charging and discharging powers to be zero when the car is not located at a station. 
Finally, called $\mathcal{V}_{l,s}$ the set of EVs located at station $s$ at time $t$,  $\mathcal{U}_s$ the rectangular box set of power limits at station s,  the last two equations \eqref{eq:s_constr_u} and \eqref{eq:s_constr_c} represent the station constraints on maximum power and available number of charging stations, respectively. 
The problem composed by equations \ref{eq:battery_of} -  \ref{eq:s_constr_c} is very general, however it is computationally expensive; due to the presence of the soft constraint on the minimum required energy \eqref{eq:soft_constr} (and to the possible quadratic objectives included in $F(u)$), the problem belongs to the MIQP class, with a number of variables in the order of $O(Tn_v)$, where in our case $n_v$ is in the order of $10^3$ and $T$ is equal to 96, since we consider 15 minutes steps and a daily control horizon. We now discuss how the original problem can be simplified by relaxing or removing some of the constraints \ref{eq:u_bilin} -  \ref{eq:s_constr_c}, and the implications for the problem's formulation hypothesis. 
\paragraph*{Strictly stationary mobility model} if the sharing model is strictly stationary, meaning that the EVs are permanently assigned to a charging station and can only be plugged there, we can relax equations \eqref{eq:s_constr_u} and \eqref{eq:s_constr_c} which encode the maximum power and connection limits per station. These can be rewritten as:
\begin{align}
&\sum_{v \in \mathcal{V}_{s}} u_{c, t, v} - u_{d, t, v} \in \mathcal{U}_s  \quad  \forall t \in \mathcal{T}, s \in \mathcal{S}  \label{eq:s_constr_u_rel}\\
&\sum_{v \in \mathcal{V}_{s}} c_{t, v}  \leq  n_{max, s} \quad     \forall t \in \mathcal{T}, s \in \mathcal{S} \label{eq:s_constr_c_rel}
\end{align}
The only difference to equations \eqref{eq:s_constr_u} and \eqref{eq:s_constr_c} is that the set $\mathcal{V}_{s}$ is no more time dependent. This effectively removes the interlink between different stations given by EVs travelling between them; in other words, sets of EVs belonging to different stations will not influence each other directly, but only by means of the system-level objective $F(u)$. Since the rest of equations (\ref{eq:u_constr}) - (\ref{eq:u_bilin}) do not interlink stations, the problem can be easily decomposed. It must be noted that the original problem can also be decomposed; however, if the mobility model is not strictly stationary, it is likely that the influencing graph between EVs is dense, meaning that the behaviour of a given EV can be influenced by a high number of other EVs, dependent on the routing between stations. This will require to introduce decoupling variables for all the states and control variables, which involves a message passing of variables in the order of $O(Tn_vn_s)$ at each iteration. On the contrary, when $F(u)$ is an aggregate function, as in all the cases presented in this paper, decomposing the problem requires messages with size in the order of $O(Tn_s)$ at each iteration. Since $n_s << n_v$ and $n_v$ is in the order of thousands, the strictly stationary hypothesis will results in a data transmission reduction in the order of $10^4$.

\paragraph*{Stations are not downsized} each station has enough chargers to accommodate all its assigned EVs at the same time. This hypothesis, combined with the previous one, allows us to remove completely the binary variable $c$ indicating whether an EV is connected to a charger. In fact, equation \eqref{eq:s_constr_c} is not needed anymore, and equations \eqref{eq:u_loc} can be replaced with:
\begin{equation}\label{eq:u_loc_rel}
u_c \preccurlyeq l u_{c, max}^T \qquad u_d \preccurlyeq l u_{d, max}^T 
\end{equation}
where $l$ is the location matrix parameter, with entries $l_{t, v}$ equal to $0$ if the $v_{th}$ vehicle is not located in any stations at time $t$.

\paragraph*{EVs are mono-directional} this hypothesis will not allow to consider direct discharge of EVs into the main grid, nor energy arbitrage between EVs. Considering currently available solutions this is the setting with lower technological burden which could already be implemented by most EV car sharing providers. Note that it will be still possible to provide services to the grid by modulating the overall charge. This hypothesis will simplify the dynamics equations, removing the discharging variable $u_d$. As a result, bi-linear constraints \eqref{eq:u_bilin} can be dropped, removing the binary variable $x_c$. If this hypothesis is combined with the two previous ones, the overall problem becomes linear or quadratic, depending on the form of $F(u)$, allowing to use a larger set of solvers and substantially reducing the computational complexity.

\subsection{Decomposition and business models}\label{sec:decomposition}
In this section we show how the original problem can be decomposed by stations under the hypothesis of a strictly stationary mobility model and that stations are not downsized. As we keep the bidirectional hypothesis, we still need to include the bilinear constraint $u_c \odot u_d = 0$, handled by equations \eqref{eq:u_bilin} and by the integer variable $x_c$. In the next session we will discuss alternative methods to handle this bilinear constraint.
Under the aforementioned hypothesis the problem can be decomposed using the alternating method of multipliers (ADMM) \cite{boyd_distributed_2010}. Following the standard ADMM procedure, since we want to decompose per station, we should introduce $n_s$ auxiliary variables representing the total power at each charging station. However, since in our case we are only interested in objective computed at the aggregation level of stations or for the overall fleet, $F(u)$ can be written in the form 
\begin{equation}
    F(u) = S\left(\sum_{s \in \mathcal{S}} p_s(u)\right) + \sum_{s \in \mathcal{S}} C(p_s(u))
\end{equation} 
where $S$ is a system level objective, that is the objective to minimize at fleet level, and $C$ is a cost function that should be minimized at station level. Here  $p_s(u) = \hat{p}_{s, load} - \hat{p}_{s, pv} +\sum_{v\in\mathcal{V}_s} u_{c,v} -u_{d,v}$ is the sum of forecasted base load and PV production (if any) for station $s$ and the sum of the charging and discharging operations of all EVs belonging to $s$. Considering this form for $F(u)$, we need to introduce only one additional variable $z \in \mathds{R}^{T}$ representing the average power of the $n_s$ controlled stations. The final problem before the decomposition can be written as:
\begin{align}
&u^{*} =  \argmin{\mathcal{X}}{S(z n_s) + \sum_{s\in\mathcal{S}} C(p_s) + Q(x)} \label{eq:battery_of_2}\\ 
s.t. & \eqref{eq:state_dyn}, \eqref{eq:u_constr},  \eqref{eq:u_bilin}, \eqref{eq:u_loc_rel}, \eqref{eq:s_constr_c_rel},  \eqref{eq:s_constr_u_rel}  \\
& z = \frac{1}{n_s}\sum_{s\in \mathcal{S}} p_s(u) \coloneqq \overline{p}_s(u) \label{eq:battery_of_2_end}
\end{align}

We can then proceed to formulate the augmented Lagrangian objective function in scaled form: 
\begin{equation}
    L_{\rho} = S(z n_s) + \sum_{s\in\mathcal{S}} C(p_s) + Q(x) + \frac{\rho}{2} \Vert \overline{p}_s(u_v) - z + \lambda \Vert_2^2
\end{equation}
Since problem \eqref{eq:battery_of_2}-\eqref{eq:battery_of_2_end} can be seen as a sharing problem, we can further simplify the standard ADMM following the description in \cite{boyd_distributed_2010} for this specific case. As the choice of ADMM's parameter to achieve a good convergence rate can be problematic under the presence of equality constraints, we use a slightly different form, namely the linearized ADMM \cite{he_non-ergodic_2015,xu_admm_nodate}; briefly speaking, this form introduces a quadratic penalty for deviating from the decision actions at the previous iteration. We can then write the minimization in the primal and dual variables update as:
\begin{align}
    & u_s^{k+1} = \argmin{u_v}{C(p_s(u_s)) + Q(x_s) + \frac{\rho}{2} \Vert p_s(u_s) - r_u^k \Vert_2^2} \label{eq:station_problem}\\ 
    &\qquad\qquad\qquad  + \frac{\gamma}{2}\Vert u_s -u_s^{k}\Vert_2^2  \label{eq:dumping}\\  
    & \qquad\qquad s.t. \ \eqref{eq:state_dyn}, \eqref{eq:u_constr},  \eqref{eq:u_bilin}, \eqref{eq:u_loc_rel}, \eqref{eq:s_constr_c_rel},  \eqref{eq:s_constr_u_rel}  \label{eq:station_constr}\\
    & z^{k+1} = \argmin{z}{S(z n_s) + \frac{\rho}{2} \Vert r_z - z^k\Vert_2^2}\\
    & \lambda^{k+1} = \lambda +\overline{p}_s(u_s)^{k+1} - z^{k+1} \label{eq:station_problem_last}
\end{align}
where $u_s = [u_v^T]^T \ \forall \ v \in \mathcal{V}_s$ and $u_s = [x_v^T]^T \ \forall \ v \in \mathcal{V}_s$ are the vectors of operations and states of all the EVs belonging to station $s$. Following \cite{boyd_distributed_2010},  $r_u^k = p_s(u_s)^k - \overline{p}_s(u_s)^k + z^k -\lambda^k$ and $r_z^k = \overline{p}_s(u_s)^{k+1} +\lambda^k$ are the reference signals for the $u$ and $z$ update. Line \eqref{eq:dumping} contains the dumping term of the linearized ADMM form for the primal variable $u_s$ update, $\gamma$ being a dumping parameter. 

The two functions $C(p_s(u_s))$ and $S(zn_s)$, representing respectively the station and the fleet objectives, can be used to tackle different business models. For example, for the station level, the following cases can be easily considered:
\begin{itemize}
    \item Minimize energy costs. Called $p_{buy} \in \mathds{R}^T$ and $p_{sell} \in \mathds{R}^T$ the time-dependent buying and selling prices in $cts/kWh$.
    In the presence of local generation e.g. due to PV power plants at the station's location, the cost function can be either positive or negative, depending on the overall power at a given time and can be expressed as in equation \eqref{eq:cost_fun}. 
    \begin{equation}\label{eq:cost_fun}
    C(p_{s,t}) = 
    \begin{cases}
    p_{buy,t} p_{s,t} , & \text{if } \quad p_{s,t} \geq 0  \\
    p_{sell,t} p_{s,t} ,              & \text{otherwise}
    \end{cases} 
    \end{equation}
    
    The cost can be thought of as the maximum over two affine functions (the first and second line of equation \eqref{eq:cost_fun}, respectively).  If $p_{buy}$ is always greater than $p_{sell}$ we can minimize energy costs by introducing an auxiliary variable $y \in \mathds{R}^T$ representing the station's energy costs. We can restrict the feasible space for $y$ to the epigraph of the cost function $C(p_s(u_s))$ by adding the two following constraints to the station problem \eqref{eq:station_problem}-\eqref{eq:station_constr}:
    \begin{align}
        y &\geq p_{buy} p_s \\
        y &\geq p_{sell} p_s
    \end{align}
    Minimizing $y$ then guarantees that its value at the optimum, $y^*$, will lie on the epigraph's lower boundary (and will thus represents the prosumer's total costs). 
    In this case $C(p_s(u_s)) = \sum_t^T y_t \delta t /3600$ where $\delta t$is the considered time step. Even without setting a system-level objective, this strategy can result in some EVs performing arbitrage, charging at low price times and later discharging to other EVs if the price swing is high enough to compensate for the round-trip efficiency.   
    \item Maximize self consumption - minimize energy imports from the grid. This can be achieved setting $C(p_s(u_s)) = \sum_t p_{s,t}(u_{s,t})$. If the term $A_v$ in the dynamic state equation \eqref{eq:state_dyn} is less than one, i.e. if self-discharging is considered, this will result in a delay-charging strategy, pushing charging operations closer to EV departure times.
    \item Minimize charging times. If we want to charge EVs up to their required SOC at departure as soon as possible, we can minimize $C(p_s(u_s)) = \sum_t p_s(u_s)d(t)$ where $d(t)$ is a convex discount function weighting less initial steps. 
    \item Perform peak shaving. The most straightforward way is to set $C(p_s(u_s)) = \Vert p_s(u_s)\Vert_2^2$. However, pure peak shaving has usually no economic drivers; the fleet manager is usually interested in reducing its total costs rather than having a flat profile per-se. Since peak tariffs are usually computed on the maximum power peak attained on a monthly basis, a more appropriate approach could be to implement a lexicographic strategy, at first minimizing the station's economic costs and then using the optimal cost found in this first step as a constraint for a second optimization in which a peak shaving objective is minimized.
\end{itemize}

At the same way, the system level objective $S(zn_s)$ can be used to address several fleet-level business cases:
\begin{itemize}
    \item Intra-day cost minimization. In the case in which the fleet manager has a deal to buy energy at intra-day costs, it can follow the same strategy illustrated to the cost minimization objective at station level and set $S(zn_s) = \sum_t^T y_t \delta t /3600$. 
    \item Profile tracking. A standard quadratic profile tracking can be used to make the fleet dispatchable setting $S(zn_s) = \sum_t^T (zn_s-r)^2$, where $r$ is a reference profile to be tracked. However, to quantify revenues from grid regulation services and flexibility calls, a linear cost function is more appropriate, as equation \eqref{eq:flex} that we used in the presented case study in section \ref{sec:economic_results}. 
\end{itemize}


\subsection{Bilinear constraints handling}
We now present the proposed method to handle the bilinear constraint $u_c \odot u_d = 0$ inside the ADMM iterations of the decomposed problem \eqref{eq:station_problem}-\eqref{eq:station_problem_last}, without using the integer variable formulation encoded in equation \eqref{eq:u_bilin}. Linear complementarity constraints arise in a variety of problems from bilevel optimization to eigenvalue complementary problems. Given a scalar objective function $f(x, y)$ of two variables $x, y \in \mathds{R}^T_+$, the simplest form of the complementarity constraint problem can be written as:
\begin{align}\label{eq:complconstr}
    \argmin{z}& f(z) \\
    s.t.& \ x^Ty=0
\end{align}
where $z = [x^T, y^T]^T$. Depending on the complexity of the underlying problem, which is in general NP-hard, different iterative methods exist to find a feasible solution or a stationary point for this kind of problem \cite{judice_optimization_2014}. One of the most used strategy is the one implemented in the YALMIP package for Matlab, which uses the built-in solver for non-convex problems BMIBNB. The procedure sequentially finds refinements of an upper and a lower bounds for the problem, respectively found using a local non-linear and a convex solver. The next iteration is then found using a standard branch-and-bound logic and split the feasible space into two new boxes \cite{noauthor_envelope_2016}. The convex approximation for bilinear problems is found using a McCormick formulation. In \cite{castro_tightening_2015}, the authors proposed tighter bounds for bilinear problems exploiting McCormick relaxations and a sequence of MILP problems. The McCormick envelope has been also proposed for the relaxation of factorable functions by systematic subgradient construction \cite{mitsos_mccormick-based_2009}, a concept similar to automatic differentiation. 
In this work we have chosen a different approach relying on the following observation: since we are solving the main problem iteratively, we want to exploit an iterative relaxation running in parallel with the standard ADMM iteration, without relying on branch and bound methods. Running a partial optimization for one part of the objective function for ADMM is theoretically justified by the generalized form of ADMM (GADMM) introduced in \cite{eckstein_douglasrachford_1992}. The GADMM guarantees the convergence even in the case in which the local (stations') problems are only partially solved. This allows us to use a first order Taylor expansion around the previous solution to approximate the complementarity constraint $x \odot y=0$, in combination with a standard ADMM using Lagrangian relaxation. We can write the first order Taylor expansion around the previous solution as:
\begin{align}
\begin{split}
    \tilde{c}( z^k, z^{k-1}) = x^{k-1} y^{k-1} &+x^{k-1}(y^k -y^{k-1}) \\ \label{eq:taylor_exp} 
    &+ y^{k-1}(x^k -x^{k-1})
\end{split}
\end{align}
We propose to use this to minimize $f(z)$ while respecting the constraint, as reported in algorithm \ref{alg:taylor}. 
\begin{algorithm}
	\SetAlgoNoLine
	\LinesNumbered
	\DontPrintSemicolon
	\SetNlSty{texttt}{}{}
	\KwIn{$z_0 = [x^T, y^T]^T$, $w_0$, $\lambda$ chosen at random, parameters $\rho$, $\gamma$}  
	\While{stop condition not met}{
	$z^{k+1} \gets \argmin{z} f(z) + \frac{\rho}{2}\Vert z-\tilde{c}(z^k, z^{k-1}) +\lambda_{k}\Vert$ \;
	$w^{k+1} \gets \frac{\rho}{\rho+\gamma}(\tilde{c}(z^k,  z^{k-1})+\lambda^k)$ \;
	$\lambda^{k+1} \gets \lambda_{k} + w^{k+1} -\tilde{c}(z^k, z^{k-1})$ \;
	$z^{k+1} = \alpha z^{k+1} + (1-\alpha) z^k$
	\caption{Taylor relaxation}\label{alg:taylor}
	}
\end{algorithm}
Here $w$ is an auxiliary variable representing $x\odot y$, which we want to shrink to zero; lines 2-4 are standard ADMM iterations where line 3 is the analytical solution of the minimization of the Lagrangian function with respect to $w$; finally line 5 is a dumped iteration over the last solution, with dumping parameter $\alpha$. 
A different approach is proposed by Wang et al. in \cite{wang_global_2018}, where they provided algorithm \ref{alg:wang}, which is a standard application of ADMM to two objective functions, $f(z)$ and $\mathds{I}_{x^Ty=0}$, where $\mathds{I}_{x^Ty=0}$ is the feasible set for the complementarity constraint. Contrary to algorithm \ref{alg:taylor} that we propose, this approach guarantees that the problem always satisfies the complementary constraint at each iteration, due to the projection onto the feasible space of $\mathds{I}_{x^Ty=0}$ at line 3. The authors proved that algorithm \ref{alg:wang} converges into a stationary point for the bilinear constrained problem when $f(z)$ is a smooth function. Algorithms \ref{alg:taylor} and \ref{alg:wang} are appealing since they are easily implementable and don't require to sequentially explore the whole solution space with a branch-and-bound strategy. 

\begin{algorithm}
	\SetAlgoNoLine
	\LinesNumbered
	\DontPrintSemicolon
	\SetNlSty{texttt}{}{}
	\KwIn{$z_0 = [x^T, y^T]^T$, $\tilde{z}_0 = [\tilde{x}^T, \tilde{y}^T]^T$, $\lambda$ chosen at random, parameter $\rho$}  
	\While{stop condition not met}{
	$z^{k+1} \gets \argmin{z} f(z) + \frac{\rho}{2}\Vert z-\tilde{z}^k +\lambda_{k}\Vert$ \;
	$\tilde{z}^{k+1} \gets \mathds{\pi}_{\tilde{x}^T\tilde{y}=0}(\tilde{z}^k-\lambda^k)$ \;
	$\lambda^{k+1} \gets \lambda_{k} + z^{k+1} - \tilde{z}^{k+1}$ \;
	\caption{Wang relaxation}\label{alg:wang}
	}
\end{algorithm}

\section{Numerical simulations}

\subsection{Data analysis and preprocessing}


We test our optimization framework on a dataset made available by a car sharing operator managing a fleet of around 3000 vehicles. 
The dataset covers all car reservations from 1st of January 2019 until 31st of July 2020, thereby including the period before the COVID-19 pandemic as well as the first wave. 
In total, there are around 2 million bookings during this period, comprising 140880 unique users and 4461 vehicles. Due to the setting of the considered car sharing service, only a small fraction of trips are one-way (0.3\%), and during the observation period only 3.5\% trips involved electric vehicles. Furthermore, the number of vehicles per station is low on average in the considered system. 73\% of all stations offer a single vehicle, further 15\% only two vehicles. 5\% of all stations have five or more vehicles. The limited availability of parking slots per station also explains the low fraction of one-way trips.


\begin{figure}
    \centering 
    \includegraphics[width=0.8\columnwidth]{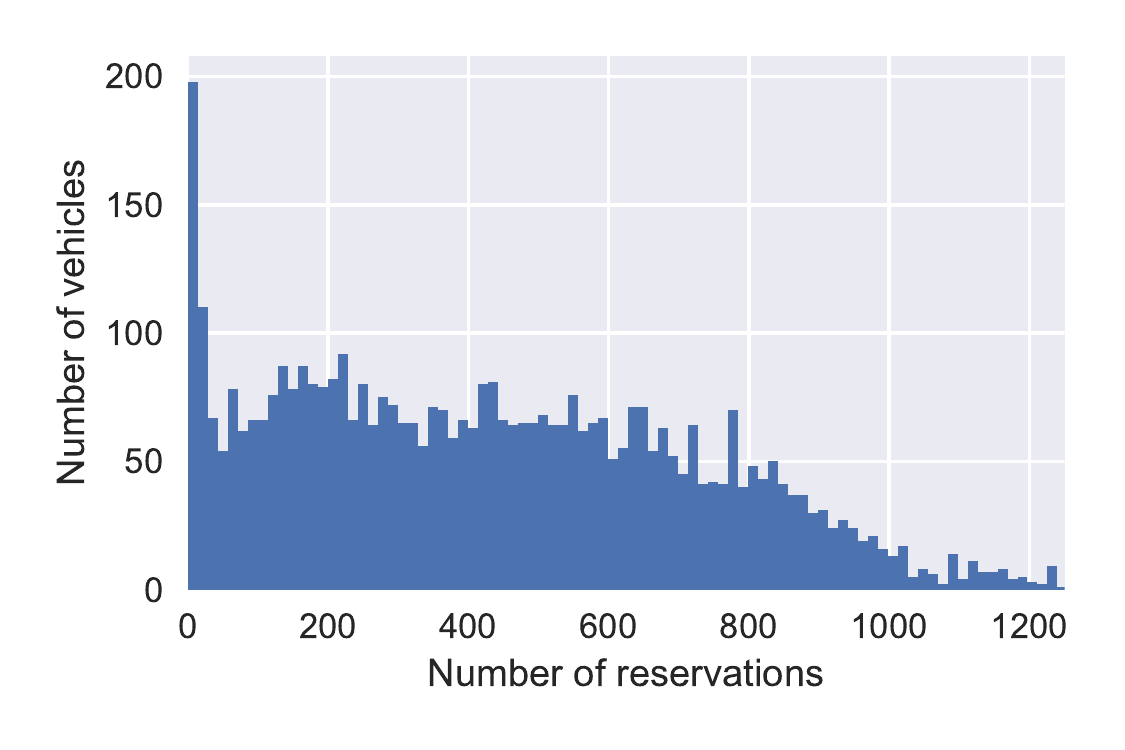}
    \caption{Reservations by vehicle}
    \label{fig:res_by_vehicle}
\end{figure}



\begin{figure}
    \centering
    \includegraphics[width=0.8\columnwidth]{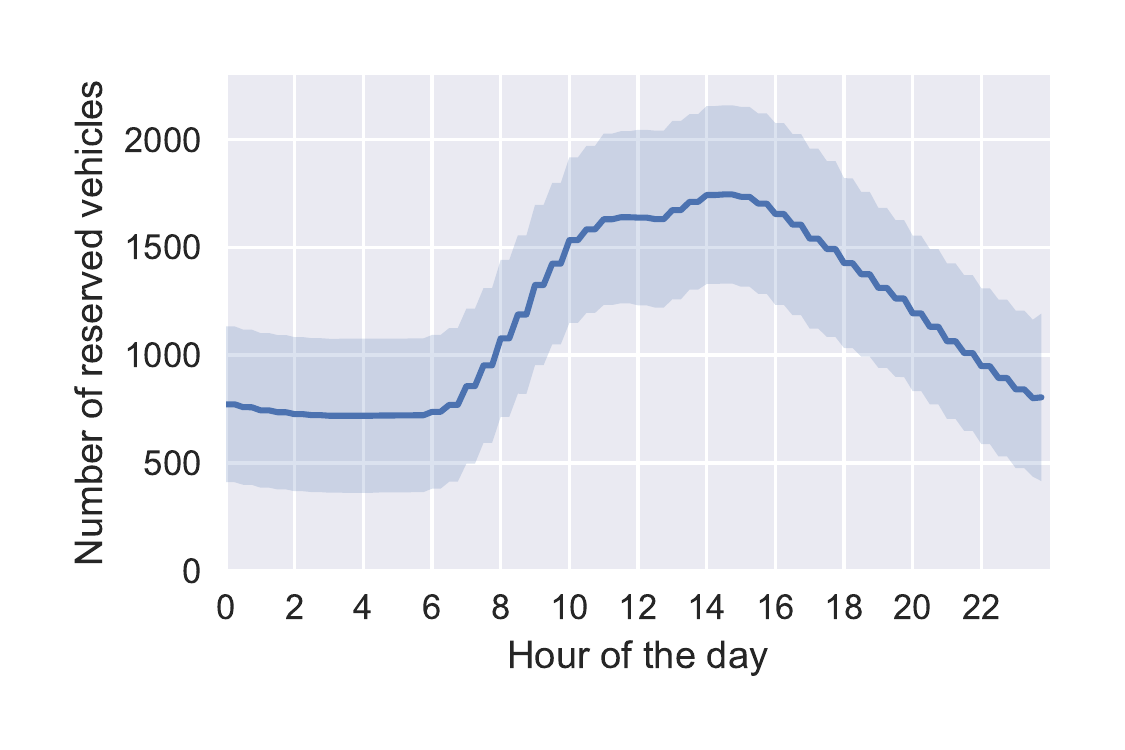}
    \caption{Reserved vehicles by time of the day}
    \label{fig:usage_by_hour}
\end{figure}


We first analyze the flexibility of vehicles for V2G operations based on their daily and overall demand. \autoref{fig:res_by_vehicle} shows the histogram of reservations by vehicle. Clearly, there are strong differences in the usage patterns of different vehicles. 48\% of the vehicles have at least one reservation in less than 50\% of the days. These findings imply a strong opportunity for the car sharing operator to utilize its fleet for V2G. However, the most flexibility is given during the night: \autoref{fig:usage_by_hour} shows a bell shaped curve of vehicle utilization over the course of a day, peaking in the afternoon. On average 21\% of vehicles are reserved at any time. Last, we validate the assumption that most car reservations are known in advance, as it is necessary for optimizing the charging schedule.  Concerning the spontaneity of the bookings, around 34\% cars are reserved more than a day in advance, whereas 20\% of the reservations are done less than an hour before the reservation period.

The data are discretized to a temporal resolution of 15-minute steps. We remove cancelled trips but include service reservations necessary for relocating vehicles.  We use the \textit{reservation} period in contrast to the actual \textit{driving period} to define the time span of car usage. However, this leads to overlapping trips in some cases when a returned vehicle was taken by the next user before the end of the original reservation period. The reservation period is therefore cut to the end of the previous drive / start of the next drive if necessary. Reservations without a ride are assumed to be cancelled and are not taken into account.

\subsection{ICE mobility patterns and State of Charge modeling}\label{sec:soc_modeling}
The car sharing service operator has set the ambitious goal to electrify their entire fleet by 2030.\ 
In order to provide a realistic simulation of the future fleet, and to demonstrate how our optimization approach scales with the number of stations, we propose to utilize the booking patterns of ICE vehicles as projected EV usage patterns, under the assumption of a similar driving behavior. Since only 3.5\% of all trips are EV trips, this scales up the number of reservations by a factor of more than 25.  
In consultation with the car sharing operator we assign an EV model to each ICE vehicle based on the car category in the car sharing operator service, i.e. "Budget", "Combi", "Transporter" etc. For example, all vehicles of the category "Transporter" were simulated as Mercedes-Benz eVito vehicles, and all in category "Budget" were assigned the VW e-up model. 

Two pieces of information are needed as input to the optimization problem: When a vehicle is plugged in at a station, and the required state of charge at the start of a reservation. Due to the modeling of ICEs as EVs and the lack of SOC data in the provided dataset, we approximate the latter by the number of driven kilometers. Given the vehicle specifications (i.e. battery range and battery capacity) we compute the required SOC by multiplying the number of driven kilometers with the average energy consumption. 

\subsection{Formulations comparison}
\begin{figure}[ht]
	\centering
	\includegraphics[width=1\columnwidth]{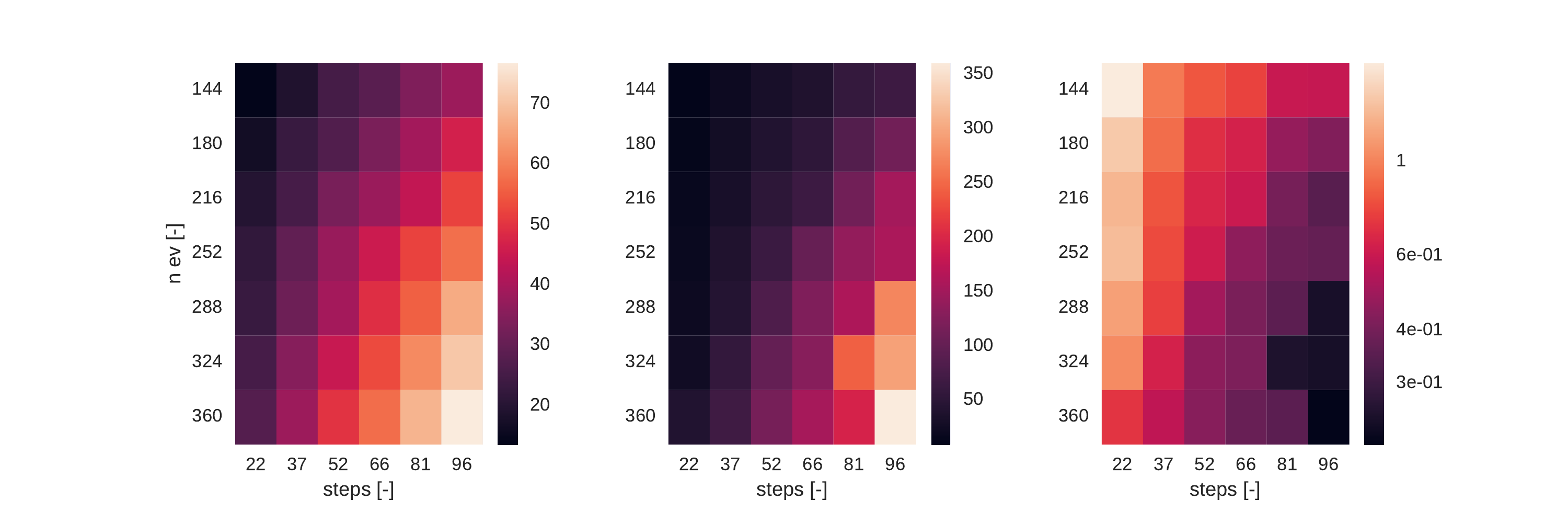}
	\caption{Computational time for different number of timesteps and considered EVs for the decomposed (left plot), the monolithic formulation (center plot) and the ratio of the two (right plot). }
	\label{fig:comp_time_monolitic}
\end{figure}

\begin{figure}[h]
	\centering
	\includegraphics[width=1\columnwidth]{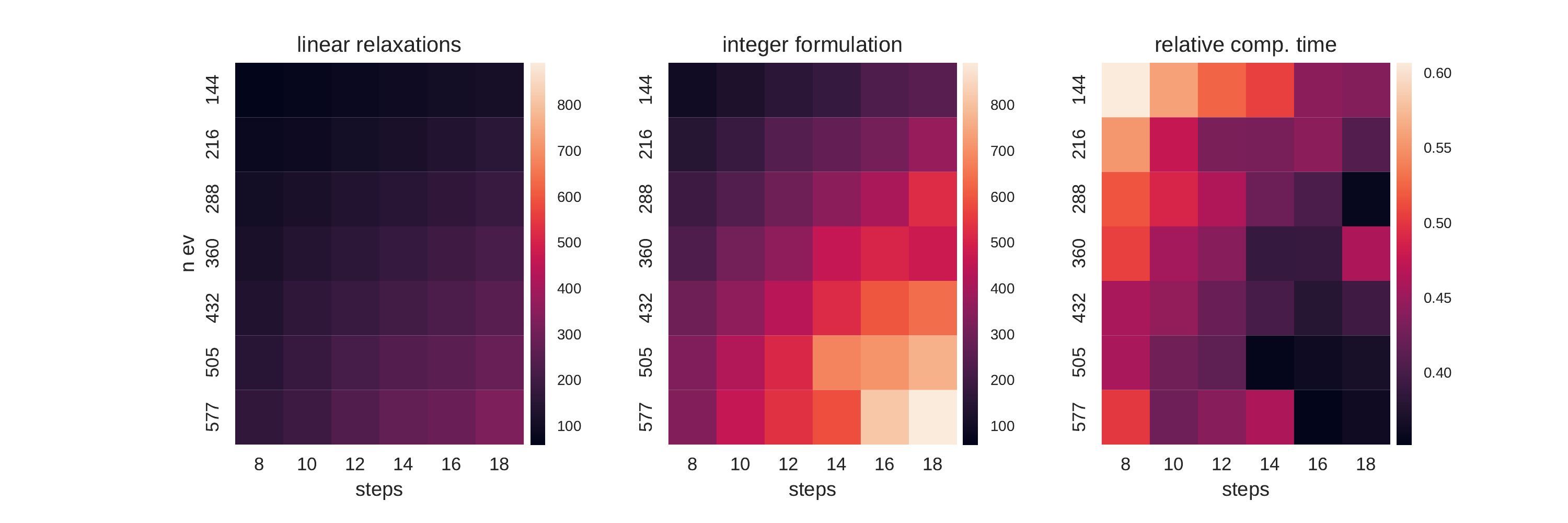}
	\caption{Computational time for different number of timesteps and considered EVs for the decomposed problem using the Taylor bilinear relaxation (left plot), the integer formulation (center plot) and the ratio of the two (right plot).}
	\label{fig:comp_time}
\end{figure}

We evaluated the numerical advantage of the proposed formulations in two steps. At first, we compared the monolithic formulation \eqref{eq:battery_of_2}-\eqref{eq:battery_of_2_end} to the decomposed one \eqref{eq:station_problem}-\eqref{eq:station_problem_last} using integer variables for handling bilinear constraints. In a second step, we evaluated the decrease in computational time in using the proposed linear methods for the bilinear constraints in the decomposed problems. For both these comparisons we vary the range of total EVs and the horizon length. The stations' objective function was set to energy cost minimization, while the system level objective was set to a profile tracking with a zero reference profile. The results of the first comparison are reported in the heatmaps of figure \ref{fig:comp_time_monolitic}. For this comparison, we solved the monolithic problem using GUROBI with standard absolute and relative tolerances, while the stopping criterion for the decomposed formulation is a joint condition on the primal and dual residual, as described in $\S$3.3.1 of \cite{boyd_distributed_2010}, using $\epsilon^{abs}=1e-6$ and $\epsilon^{rel}=1e-4$, respectively. The first two heatmaps refer to the total computational time of the decomposed problem and the monolithic formulation, respectively. The last plot shows the ratio of the two, a value lower than one meaning a lower computational time for the decomposed formulation. As expected, the computational advantage over the monolithic formulation increases with both the number of EVs and the length of the horizon. The experimental data for up to 360 vehicles shows a clear trend; the computational time of the decomposed problem for the most time consuming configuration being roughly 20\% of the time needed by the monolithic formulation. The second comparison was done using a fixed number of iterations, which was set to 800. At first, we tuned the parameters of algorithm \ref{alg:taylor} and \ref{alg:wang} w.r.t. the solution reached by the integer formulation, using a random sampling strategy over the configuration with 144 EVs and an 18 steps horizon. The parameters ($\rho$ and $\gamma$ for \ref{alg:taylor} and $\rho$ for \ref{alg:wang}, respectively) were then held constant over the different combinations of EVs and horizon lengths. We found that both the algorithms' performance was stable for a large range of parameters values. The computational times are shown in figure \ref{fig:comp_time}, where the first heatmap refers to the Taylor relaxation, the second one to the integer formulation and the last is the ratio of the two. As the computational advantage is due to the change of the class of the problem from MIQP to QP, we found a negligible difference in the computation times between algorithm \ref{alg:taylor} and \ref{alg:wang}, and thus here report only results for the Taylor relaxation. Also in this case there is a clear trend in the reduction of computational time with increasing number of EVs and steps. The highest reduction was found for the most time consuming configuration of 577 EVs and 18 steps, with the Taylor relaxation using roughly 35\% of the time needed by the integer formulation; once again we expect this value to get lower for problems with higher number of EVs.   

\begin{figure}[h]
	\centering
	\includegraphics[width=0.8\columnwidth]{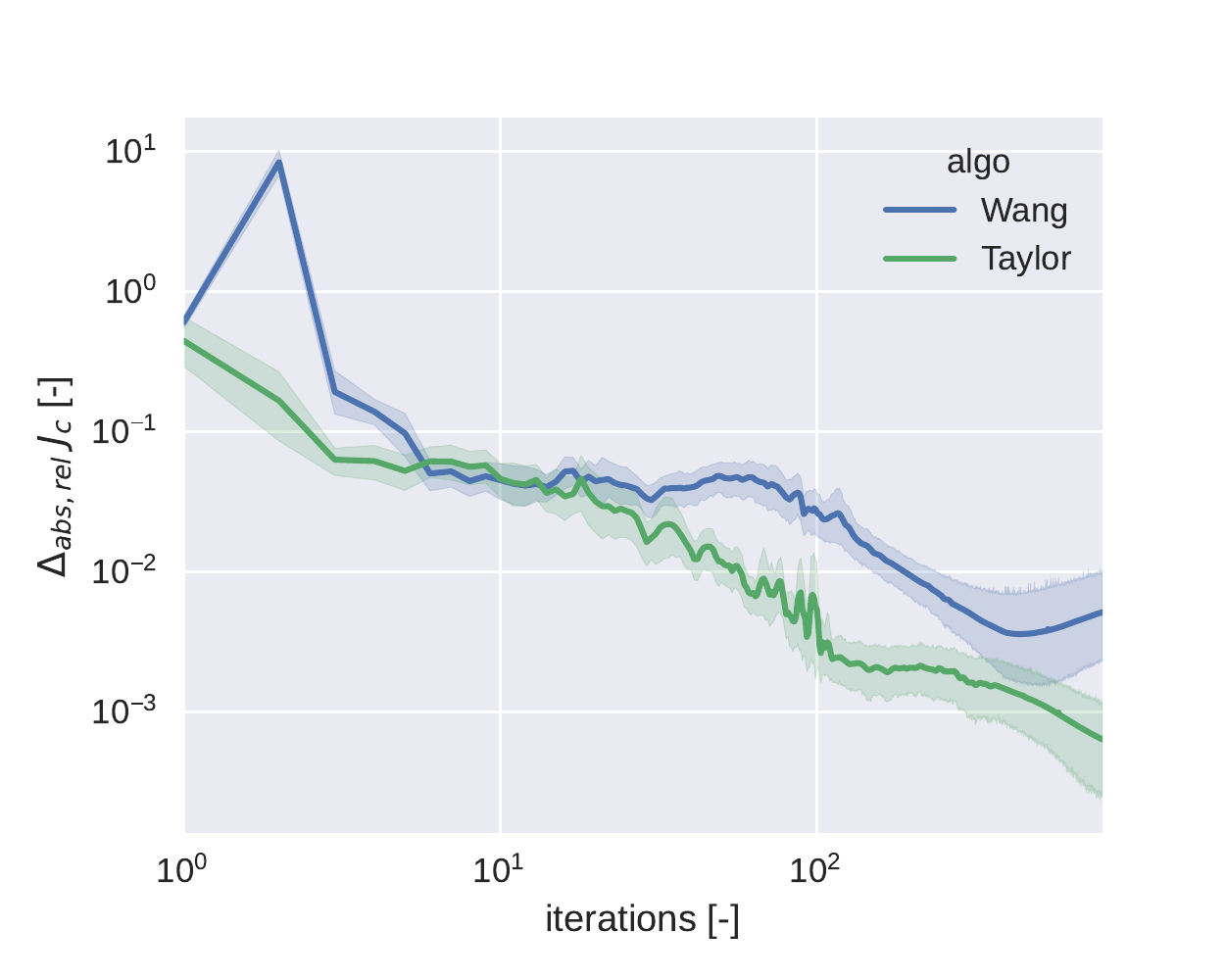}
	\caption{Comparison of convergence dynamics using the Taylor or Wang formulation for the bilinear constraint relaxation, in terms of relative differences in total objective w.r.t. integer formulation, when stations optimize for costs and the fleet has a reference tracking objective. Confidence interval refers to all the 42 combinations of horizon length and number of EVs of figure \ref{fig:comp_time}.}
	\label{fig:comp_convergence_ev}
\end{figure}

Figure \ref{fig:comp_convergence_ev} shows the distribution of $\Delta_{abs, rel} J_c$ for all the cases reported in figure \ref{fig:comp_time}. Here $J_c$ is defined as the sum of the different objective functions without including any augmented Lagrangian terms (neither the one deriving by the problem decomposition nor the ones of the linear formulations) in order to have a fair comparison:
\begin{equation}\label{eq:Jc}
    J_c = F(u) + Q(x) + \frac{\gamma}{2}\Vert u -u^{k}\Vert_2^2 
\end{equation}
Both the algorithms converge to the solution of the integer formulation with some oscillations, even if the Taylor-based relaxation shows better convergence, achieving a relative difference in the order of $1e-3$ for all the cases after 800 iterations. 

\subsection{Economic results}\label{sec:economic_results}
\begin{figure}
    \centering
    \includegraphics[width=0.8\columnwidth]{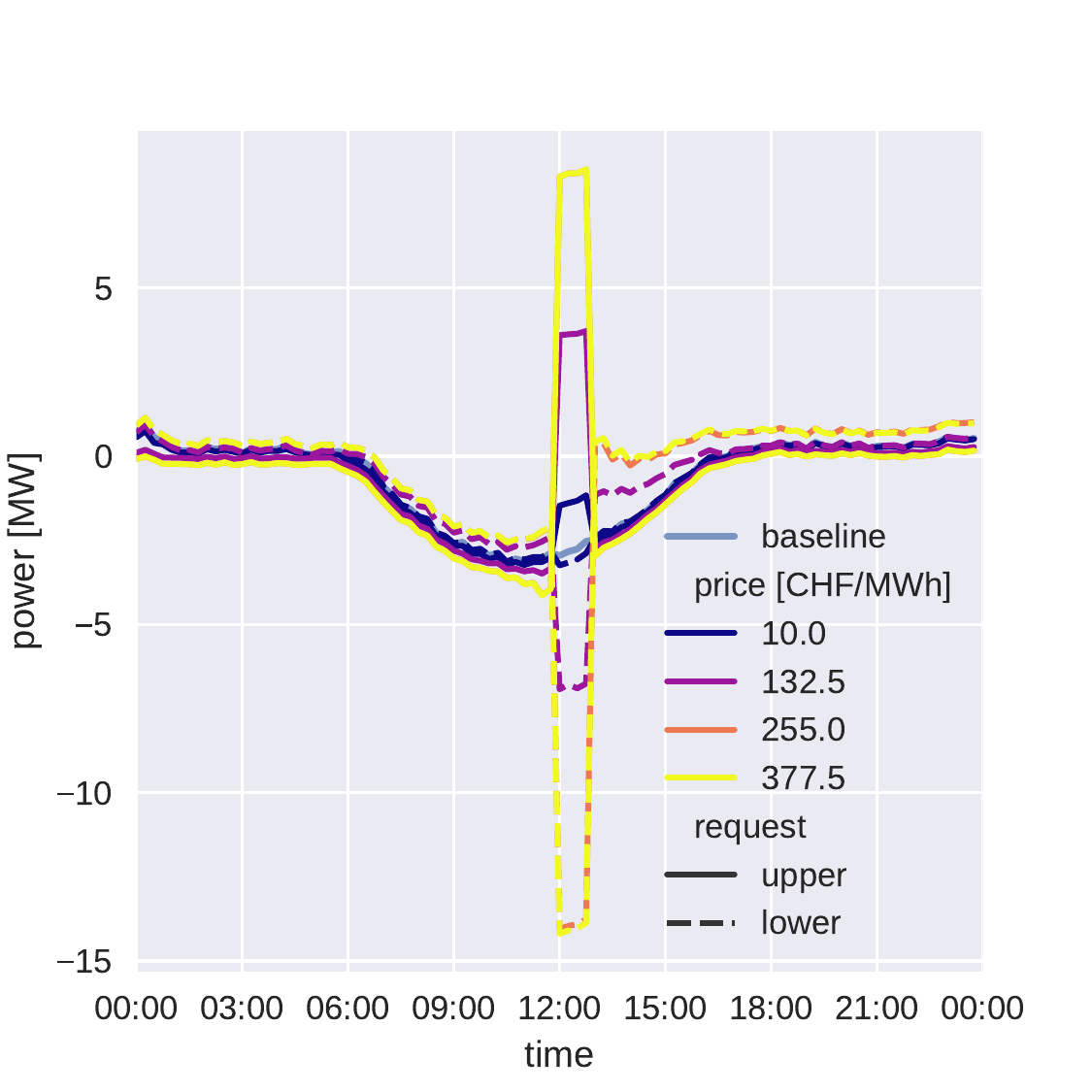}
    \caption{Example of response to upward and downward flexibility calls as a function of price, compared to the baseline case in which there is no system level costs and the stations just optimize for their local energy prices.}
    \label{fig:flexi_example}
\end{figure}

\begin{figure}
    \centering
    \includegraphics[width=0.8\columnwidth]{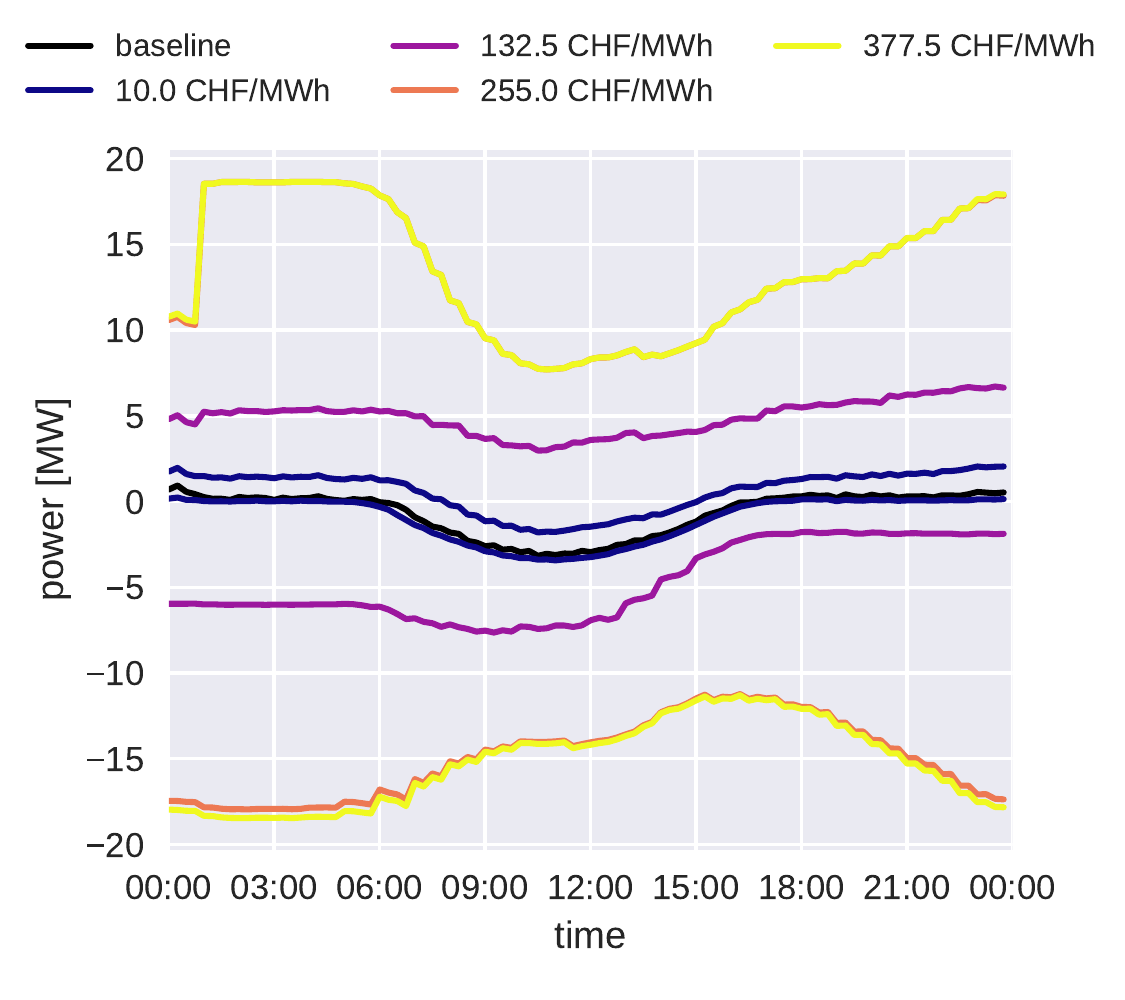}
    \caption{Flexibility envelope for different levels of $p_f$, showing the maximum attainable flexibility for hourly slots of the day.}
    \label{fig:flexi_envelope}
\end{figure}

\begin{figure}
    \centering
    \includegraphics[width=0.8\columnwidth]{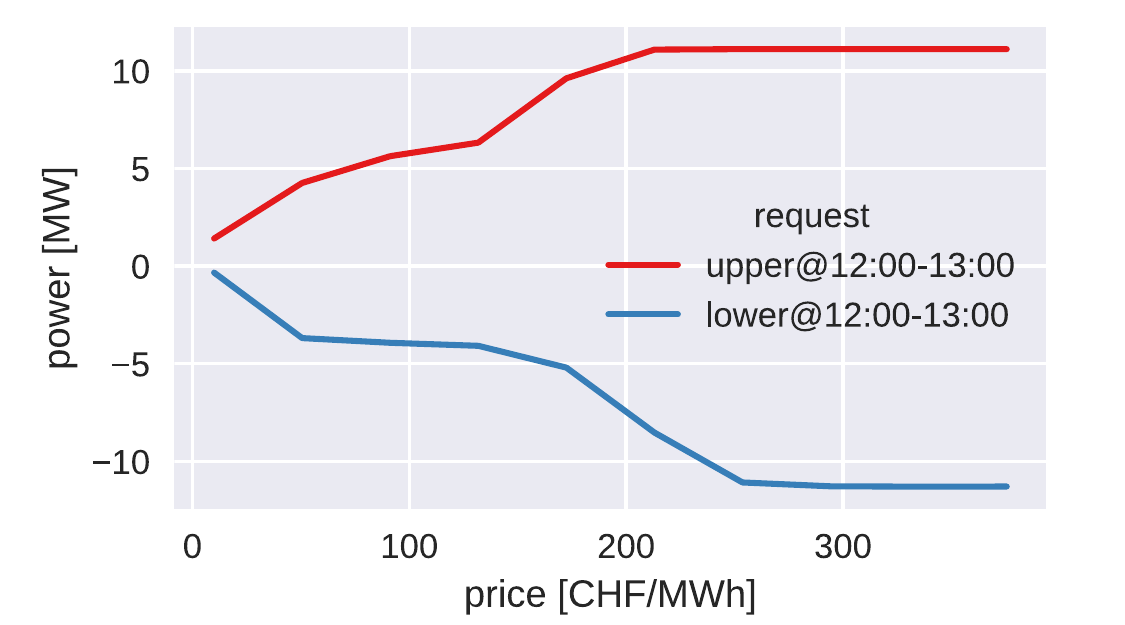}
    \caption{Deviation from the baseline power profile as a function of flexibility price $p_f$, for the noon case.}
    \label{fig:price_sigmoid}
\end{figure}

\begin{figure}
    \centering
    \includegraphics[width=0.8\columnwidth]{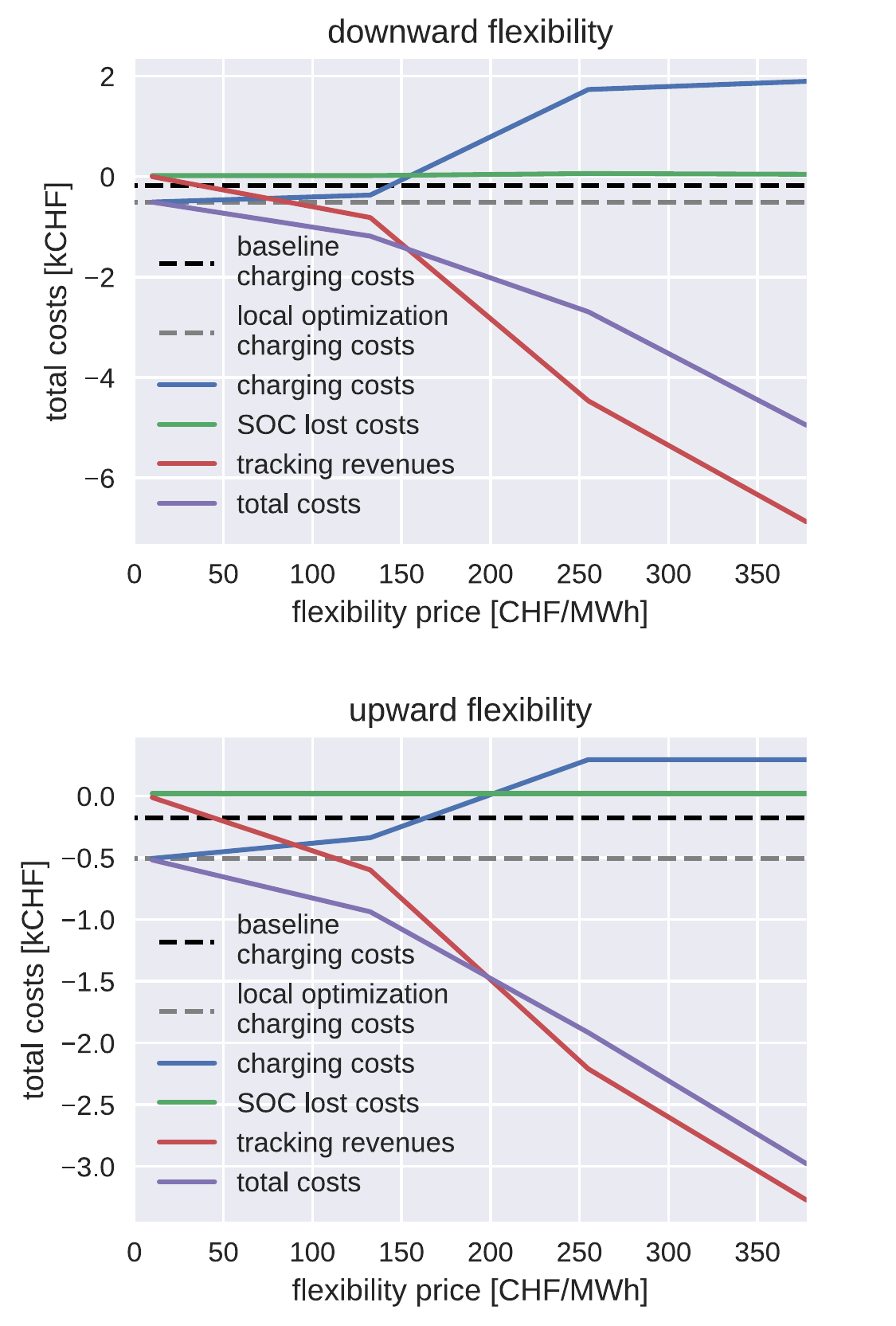}
    \caption{Behaviour of different fleet costs as a function of flexibility price $p_f$, for the noon case.}
    \label{fig:noon_costs}
\end{figure}

We use the proposed algorithm \ref{alg:taylor} to retrieve flexibility boundaries for an EV fleet. The setting is the following: an EV manager bidding for ancillary services is interested to know for a given leading time how many MWs, for how long, can be requested to the EV fleet for both upward and downward flexibility calls, and how much it costs per MWh. This information can then be used by the manager to make more informative bids. We followed the approach proposed in \cite{oldewurtel_towards_2013} to achieve hourly flexibility boundary for an aggregation of office buildings. For each hour of the day, we solve the optimization problem \eqref{eq:station_problem}-\eqref{eq:station_problem_last}, where each station minimizes its total energy costs for the EVs charging operations, and $C(p_s(u_s))$ is modeled through the auxiliary variable $y$ as explained in section \ref{sec:decomposition}. Since the considered car sharing operator's stations are located under different Swiss DSOs, we used data from \cite{elcom_basic_nodate} to link them with the correct values for the buying and selling energy prices $p_{buy}$ and $p_{sell}$, depending on their location. Additionally, we probabilistically assigned each station with a PV power plant, with a nominal power proportional to the maximum number of hosted EVs at that station. 
The system level objective function is set to be:
\begin{equation}\label{eq:flex}
    S\left(\sum_{s\in\mathcal{S}}p_s(u)\right) = \sum_{t\in\mathcal{T}_h} p_{f} \vert r-\sum_{s\in\mathcal{S}}p_s(u)\vert
\end{equation}
where $r$ is the reference profile, $\mathcal{T}_h$ is the set of timesteps belonging to hour $h$ and $p_{f}$ is the price of flexibility, which is constant over the considered hour. Equation \eqref{eq:flex} can be seen as a linear punishment in deviating from a flexibility call. We simulated a total number of 1440 EVs, keeping all the EVs belonging to a given station if the latter was chosen by a random sampling among all the available ones.
An example of results using this objective function when $h=12$, at different price levels, is shown in figure \ref{fig:flexi_example}. When the fleet receives an upward flexibility call at noon, the consumption decreases in the rest of the day w.r.t. the baseline profile in which the system level objective is set to zero and the only objective is the stations' cost minimization. The opposite verifies when the fleet receives a downward flexibility call. For a given day, we run 24 optimizations, systematically changing $\mathcal{T}_h$ and repeat the process for different values of $p_{f}$. The resulting flexibility envelopes can be seen in figure \ref{fig:flexi_envelope}. Lines of different colors represent the convex envelopes of the maximum and minimum flexibility attained at different hour of the days for a given price $p_f$. It can be seen how during the first hour of the day the fleet is not prepared to an upward call, since the average SOC of the fleet is too high and the fleet has no time to discharge beforehand. Moreover, a saturation effect can be noticed after a given level of price: the maximum attainable flexibility does not change significantly passing from a $p_f$ of 255 CHF/MWh to 377 CHF/MWh. In order to better analyze this effect, we considered more price levels for the case in which flexibility is requested at noon. Figure \ref{fig:price_sigmoid} shows the maximum amount of MW reached for 10 different values of $p_f$ ranging from 10 CHF/MWh to 377.5 CHF/MWh. The saturation effect is clear for both the upper and lower requests, but it's starting at slightly different price levels, around 210 and 250 CHF/MWh, respectively.
Finally, we study the effect of the flexibility request on the other considered costs in the optimization problem. Figure \ref{fig:noon_costs} shows the change in charging costs, loss of SOC (equation \eqref{eq:soft_constr}), tracking revenues and total costs for the noon case. As expected, as the price level increases, the tracking revenues rises for both upward and downward flexibility calls, but this comes at the expense of higher charging costs. The change of cost for the SOC lost is negligible compared to the other costs. 

\section{Conclusions}
In this paper we presented an optimization model to control the charging and discharging operation of large EV fleets. We started by modeling a generic case in which the EVs are allowed to relocate between stations, and then focused on the strictly stationary model where EVs are picked up and dropped off at the same station, since this reflects the conditions of the presented case study. For this last case we demonstrated how the problem can be decomposed by stations, allowing to reduce the overall computational time. Furthermore, we used iterative methods to handle the bilinear constraints arising from the V2G formulation, which allows us to use a larger class of (free) solvers. For different combinations of horizon's lengths and number of EVs, we reported numerical results showing substantial speed ups w.r.t. the monolithic formulation, due to both problem decomposition and the use of relaxations for the bilinear constraints. We see multiple opportunities for future work. First, 
many car sharing bookings are spontaneous, limiting the applicability of day-ahead planning in real world scenarios. This could be tackled with the integration of booking forecasts; since forecasts introduce uncertainty, a receding horizon optimization can be used to minimize errors. Additionally, a stochastic formulation e.g. tree-based stochastic MPC \cite{5399917}, can be used to further tackle the uncertainty of bookings and PV generation.  

\section*{Acknowledgments}
This work was financially supported of the Swiss Federal Office of Energy (V2G4CarSharing and GAMES projects SI/502344, SI/502361).

\bibliographystyle{abbrvnat}
\bibliography{references}

\begin{thebibliography}{39}
\providecommand{\natexlab}[1]{#1}
\providecommand{\url}[1]{\texttt{#1}}
\expandafter\ifx\csname urlstyle\endcsname\relax
  \providecommand{\doi}[1]{doi: #1}\else
  \providecommand{\doi}{doi: \begingroup \urlstyle{rm}\Url}\fi

\bibitem[noa(2022)]{noauthor_envelope_2016}
Envelope approximations for global optimization,
  https://yalmip.github.io/tutorial/envelopesinbmibnb, May 2022.

\bibitem[Bernardini and Bemporad(2009)]{5399917}
D.~Bernardini and A.~Bemporad.
\newblock Scenario-based model predictive control of stochastic constrained
  linear systems.
\newblock In \emph{Proceedings of the 48h IEEE Conference on Decision and
  Control (CDC) held jointly with 2009 28th Chinese Control Conference}, pages
  6333--6338, 2009.
\newblock \doi{10.1109/CDC.2009.5399917}.

\bibitem[Biondi et~al.(2016)Biondi, Boldrini, and Bruno]{biondi2016optimal}
E.~Biondi, C.~Boldrini, and R.~Bruno.
\newblock Optimal charging of electric vehicle fleets for a car sharing system
  with power sharing.
\newblock In \emph{2016 IEEE International Energy Conference (ENERGYCON)},
  pages 1--6. IEEE, 2016.

\bibitem[Boyd(2010)]{boyd_distributed_2010}
S.~Boyd.
\newblock Distributed {Optimization} and {Statistical} {Learning} via the
  {Alternating} {Direction} {Method} of {Multipliers}.
\newblock \emph{Foundations and Trends® in Machine Learning}, 3\penalty0
  (1):\penalty0 1--122, 2010.

\bibitem[Caggiani et~al.(2021)Caggiani, Prencipe, and
  Ottomanelli]{caggiani_static_2021}
L.~Caggiani, L.~P. Prencipe, and M.~Ottomanelli.
\newblock A static relocation strategy for electric car-sharing systems in a
  vehicle-to-grid framework.
\newblock \emph{Transportation Letters}, 13\penalty0 (3):\penalty0 219--228,
  Mar. 2021.

\bibitem[Castro(2015)]{castro_tightening_2015}
P.~M. Castro.
\newblock Tightening piecewise {McCormick} relaxations for bilinear problems.
\newblock \emph{Computers \& Chemical Engineering}, 72:\penalty0 300--311, Jan.
  2015.

\bibitem[Crozier et~al.(2020)Crozier, Morstyn, and
  McCulloch]{crozier2020opportunity}
C.~Crozier, T.~Morstyn, and M.~McCulloch.
\newblock The opportunity for smart charging to mitigate the impact of electric
  vehicles on transmission and distribution systems.
\newblock \emph{Applied Energy}, 268:\penalty0 114973, 2020.

\bibitem[Dang et~al.(2019)Dang, Wu, and Boulet]{dang_q-learning_2019}
Q.~Dang, D.~Wu, and B.~Boulet.
\newblock A {Q}-{Learning} {Based} {Charging} {Scheduling} {Scheme} for
  {Electric} {Vehicles}.
\newblock In \emph{2019 {IEEE} {Transportation} {Electrification} {Conference}
  and {Expo} ({ITEC})}, pages 1--5, June 2019.

\bibitem[Eckstein and Bertsekas(1992)]{eckstein_douglasrachford_1992}
J.~Eckstein and D.~P. Bertsekas.
\newblock On the {Douglas}—{Rachford} splitting method and the proximal point
  algorithm for maximal monotone operators.
\newblock \emph{Mathematical Programming}, 55\penalty0 (1-3):\penalty0
  293--318, Apr. 1992.

\bibitem[ElCom()]{elcom_basic_nodate}
F.~E.~C. ElCom.
\newblock Basic data for tariffs of the {Swiss} {Distribution} {Network}
  {Operators}.

\bibitem[Fournier et~al.(2014)Fournier, Lindenlauf, Baumann, Seign, and
  Weil]{fournier2014carsharing}
G.~Fournier, F.~Lindenlauf, M.~Baumann, R.~Seign, and M.~Weil.
\newblock Carsharing with electric vehicles and vehicle-to-grid: a future
  business model?
\newblock In \emph{Radikale Innovationen in der Mobilit{\"a}t}, pages 63--79.
  Springer, 2014.

\bibitem[García-Villalobos et~al.(2014)García-Villalobos, Zamora,
  San~Martín, Asensio, and Aperribay]{garcia-villalobos_plug-electric_2014}
J.~García-Villalobos, I.~Zamora, J.~I. San~Martín, F.~J. Asensio, and
  V.~Aperribay.
\newblock Plug-in electric vehicles in electric distribution networks: {A}
  review of smart charging approaches.
\newblock \emph{Renewable and Sustainable Energy Reviews}, 38:\penalty0
  717--731, Oct. 2014.

\bibitem[Garifi et~al.(2019)Garifi, Baker, Christensen, and
  Touri]{garifi_control_2019}
K.~Garifi, K.~Baker, D.~Christensen, and B.~Touri.
\newblock Control of {Energy} {Storage} in {Home} {Energy} {Management}
  {Systems}: {Non}-{Simultaneous} {Charging} and {Discharging} {Guarantees}.
\newblock \emph{arXiv:1805.00100 [math]}, Apr. 2019.

\bibitem[He and Yuan(2015)]{he_non-ergodic_2015}
B.~He and X.~Yuan.
\newblock On non-ergodic convergence rate of {Douglas}–{Rachford} alternating
  direction method of multipliers.
\newblock \emph{Numerische Mathematik}, 130\penalty0 (3):\penalty0 567--577,
  July 2015.

\bibitem[He et~al.(2021)He, Ma, Qi, and Wang]{he2021charging}
L.~He, G.~Ma, W.~Qi, and X.~Wang.
\newblock Charging an electric vehicle-sharing fleet.
\newblock \emph{Manufacturing \& Service Operations Management}, 23\penalty0
  (2):\penalty0 471--487, 2021.

\bibitem[(IEA)(2021)]{iae}
I.~E.~A. (IEA).
\newblock {Global EV Outlook 2021}.
\newblock Technical report, 2021.
\newblock URL \url{https://www.iea.org/reports/global-ev-outlook-2021}.

\bibitem[Júdice(2014)]{judice_optimization_2014}
J.~Júdice.
\newblock Optimization with linear complementarity constraints.
\newblock \emph{Pesquisa Operacional}, 34\penalty0 (3):\penalty0 559--584, Dec.
  2014.

\bibitem[Kara et~al.(2015)Kara, Macdonald, Black, Bérges, Hug, and
  Kiliccote]{kara_estimating_2015}
E.~C. Kara, J.~S. Macdonald, D.~Black, M.~Bérges, G.~Hug, and S.~Kiliccote.
\newblock Estimating the benefits of electric vehicle smart charging at
  non-residential locations: {A} data-driven approach.
\newblock \emph{Applied Energy}, 155:\penalty0 515--525, Oct. 2015.

\bibitem[Kempton and Tomi{\'c}(2005)]{kempton2005vehicle}
W.~Kempton and J.~Tomi{\'c}.
\newblock Vehicle-to-grid power implementation: From stabilizing the grid to
  supporting large-scale renewable energy.
\newblock \emph{Journal of power sources}, 144\penalty0 (1):\penalty0 280--294,
  2005.

\bibitem[Li et~al.(2019)Li, Wan, and He]{li2019constrained}
H.~Li, Z.~Wan, and H.~He.
\newblock Constrained ev charging scheduling based on safe deep reinforcement
  learning.
\newblock \emph{IEEE Transactions on Smart Grid}, 11\penalty0 (3):\penalty0
  2427--2439, 2019.

\bibitem[Ma et~al.(2016)Ma, Zou, Ran, Shi, and
  Hiskens]{maEfficientDecentralizedCoordination2016}
Z.~Ma, S.~Zou, L.~Ran, X.~Shi, and I.~A. Hiskens.
\newblock Efficient decentralized coordination of large-scale plug-in electric
  vehicle charging.
\newblock \emph{Automatica}, 69:\penalty0 35--47, July 2016.
\newblock ISSN 00051098.
\newblock \doi{10.1016/j.automatica.2016.01.035}.

\bibitem[Martin et~al.(2022)Martin, Buffat, Bucher, Hamper, and
  Raubal]{martin2022using}
H.~Martin, R.~Buffat, D.~Bucher, J.~Hamper, and M.~Raubal.
\newblock Using rooftop photovoltaic generation to cover individual electric
  vehicle demand—a detailed case study.
\newblock \emph{Renewable and Sustainable Energy Reviews}, 157:\penalty0
  111969, 2022.

\bibitem[Mitsos et~al.(2009)Mitsos, Chachuat, and
  Barton]{mitsos_mccormick-based_2009}
A.~Mitsos, B.~Chachuat, and P.~I. Barton.
\newblock {McCormick}-{Based} {Relaxations} of {Algorithms}.
\newblock \emph{SIAM Journal on Optimization}, 20\penalty0 (2):\penalty0
  573--601, Jan. 2009.

\bibitem[Oldewurtel et~al.(2013)Oldewurtel, Sturzenegger, Andersson, Morari,
  and Smith]{oldewurtel_towards_2013}
F.~Oldewurtel, D.~Sturzenegger, G.~Andersson, M.~Morari, and R.~S. Smith.
\newblock Towards a standardized building assessment for demand response.
\newblock In \emph{52nd {IEEE} {Conference} on {Decision} and {Control}}, pages
  7083--7088, Dec. 2013.

\bibitem[Ravi and Aziz(2022)]{raviUtilizationElectricVehicles2022}
S.~S. Ravi and M.~Aziz.
\newblock Utilization of {{Electric Vehicles}} for {{Vehicle-to-Grid
  Services}}: {{Progress}} and {{Perspectives}}.
\newblock \emph{Energies}, 15\penalty0 (2):\penalty0 589, Jan. 2022.
\newblock ISSN 1996-1073.
\newblock \doi{10.3390/en15020589}.

\bibitem[Sadeghianpourhamami et~al.(2019)Sadeghianpourhamami, Deleu, and
  Develder]{sadeghianpourhamami2019definition}
N.~Sadeghianpourhamami, J.~Deleu, and C.~Develder.
\newblock Definition and evaluation of model-free coordination of electrical
  vehicle charging with reinforcement learning.
\newblock \emph{IEEE Transactions on Smart Grid}, 11\penalty0 (1):\penalty0
  203--214, 2019.

\bibitem[Shaheen and Cohen(2020)]{shaheen2020innovative}
S.~Shaheen and A.~Cohen.
\newblock Innovative mobility: Carsharing outlook carsharing market overview,
  analysis, and trends.
\newblock 2020.

\bibitem[Shieh et~al.(1980)Shieh, Wang, and
  Yates]{shieh_discrete-continuous_1980}
L.~S. Shieh, H.~Wang, and R.~E. Yates.
\newblock Discrete-continuous model conversion.
\newblock \emph{Applied Mathematical Modelling}, 4\penalty0 (6):\penalty0
  449--455, Dec. 1980.

\bibitem[Tan et~al.(2016)Tan, Ramachandaramurthy, and
  Yong]{tan_integration_2016}
K.~M. Tan, V.~K. Ramachandaramurthy, and J.~Y. Yong.
\newblock Integration of electric vehicles in smart grid: {A} review on vehicle
  to grid technologies and optimization techniques.
\newblock \emph{Renewable and Sustainable Energy Reviews}, 53:\penalty0
  720--732, Jan. 2016.

\bibitem[Tuchnitz et~al.(2021)Tuchnitz, Ebell, Schlund, and
  Pruckner]{tuchnitz2021development}
F.~Tuchnitz, N.~Ebell, J.~Schlund, and M.~Pruckner.
\newblock Development and evaluation of a smart charging strategy for an
  electric vehicle fleet based on reinforcement learning.
\newblock \emph{Applied Energy}, 285:\penalty0 116382, 2021.

\bibitem[Valogianni et~al.(2013)Valogianni, Ketter, and
  Collins]{valogianni2013smart}
K.~Valogianni, W.~Ketter, and J.~Collins.
\newblock Smart charging of electric vehicles using reinforcement learning.
\newblock In \emph{Workshops at the Twenty-Seventh AAAI Conference on
  Artificial Intelligence}, 2013.

\bibitem[Wan et~al.(2018)Wan, Li, He, and Prokhorov]{wan2018model}
Z.~Wan, H.~Li, H.~He, and D.~Prokhorov.
\newblock Model-free real-time ev charging scheduling based on deep
  reinforcement learning.
\newblock \emph{IEEE Transactions on Smart Grid}, 10\penalty0 (5):\penalty0
  5246--5257, 2018.

\bibitem[Wang et~al.(2018)Wang, Yin, and Zeng]{wang_global_2018}
Y.~Wang, W.~Yin, and J.~Zeng.
\newblock Global {Convergence} of {ADMM} in {Nonconvex} {Nonsmooth}
  {Optimization}, May 2018.

\bibitem[Xu et~al.(2021)Xu, Wu, and Tan]{xu_electric_2021}
M.~Xu, T.~Wu, and Z.~Tan.
\newblock Electric vehicle fleet size for carsharing services considering
  on-demand charging strategy and battery degradation.
\newblock \emph{Transportation Research Part C: Emerging Technologies},
  127:\penalty0 103146, June 2021.

\bibitem[Xu et~al.()Xu, Liu, Lin, and Yang]{xu_admm_nodate}
Y.~Xu, M.~Liu, Q.~Lin, and T.~Yang.
\newblock {ADMM} without a {Fixed} {Penalty} {Parameter}: {Faster}
  {Convergence} with {New} {Adaptive} {Penalization}.
\newblock page~11.

\bibitem[Xu et~al.(2018)Xu, {\c{C}}olak, Kara, Moura, and
  Gonz{\'a}lez]{xu2018planning}
Y.~Xu, S.~{\c{C}}olak, E.~C. Kara, S.~J. Moura, and M.~C. Gonz{\'a}lez.
\newblock Planning for electric vehicle needs by coupling charging profiles
  with urban mobility.
\newblock \emph{Nature Energy}, 3\penalty0 (6):\penalty0 484--493, 2018.

\bibitem[Yi et~al.(2020)Yi, Scoffield, Smart, Meintz, Jun, Mohanpurkar, and
  Medam]{yiHighlyEfficientControl2020}
Z.~Yi, D.~Scoffield, J.~Smart, A.~Meintz, M.~Jun, M.~Mohanpurkar, and A.~Medam.
\newblock A highly efficient control framework for centralized residential
  charging coordination of large electric vehicle populations.
\newblock \emph{International Journal of Electrical Power \& Energy Systems},
  117:\penalty0 105661, May 2020.
\newblock ISSN 01420615.

\bibitem[Zhang et~al.(2021)Zhang, Lu, and Shen]{zhang2021values}
Y.~Zhang, M.~Lu, and S.~Shen.
\newblock On the values of vehicle-to-grid electricity selling in electric
  vehicle sharing.
\newblock \emph{Manufacturing \& Service Operations Management}, 23\penalty0
  (2):\penalty0 488--507, 2021.

\bibitem[Zhong et~al.(2014)Zhong, He, Li, Cao, Wang, Fang, Zeng, and
  Xiao]{zhongCoordinatedControlLargescale2014}
J.~Zhong, L.~He, C.~Li, Y.~Cao, J.~Wang, B.~Fang, L.~Zeng, and G.~Xiao.
\newblock Coordinated control for large-scale {{EV}} charging facilities and
  energy storage devices participating in frequency regulation.
\newblock \emph{Applied Energy}, 123:\penalty0 253--262, June 2014.
\newblock ISSN 03062619.
\newblock \doi{10.1016/j.apenergy.2014.02.074}.

\end{thebibliography}

\newpage


 




\vfill

\end{document}